\documentclass[transmag]{IEEEtran}
\usepackage{latexsym}
\usepackage{graphicx}
\usepackage{amsfonts,amssymb,amsmath}
\usepackage{hyperref}
\usepackage{graphicx}
\usepackage{gensymb}
\usepackage{color}
\usepackage{epstopdf}

\newcommand{\quotes}[1]{``#1''}
\usepackage[
left=1.55cm,
right=1.55cm,
top=1.4cm,
bottom=1.4cm
]{geometry}

\def\BibTeX{{\rm B\kern-.05em{\sc i\kern-.025em b}\kern-.08em T\kern-.1667em\lower.7ex\hbox{E}\kern-.125emX}}
\begin{document}

\title{A Cognitive Radio Enabled RF/FSO Communication Model for Aerial Relay Networks: Possible Configurations and Opportunities}

\author{Eylem Erdogan, \IEEEmembership{Member, IEEE}, Ibrahim Altunbas, \IEEEmembership{Senior Member, IEEE}, \\Nihat Kabaoglu, and Halim Yanikomeroglu, \IEEEmembership{Fellow, IEEE} 
\thanks{E. Erdogan and N. Kabaoglu are with the Department of Electrical and Electronics Engineering, Istanbul Medeniyet University, Istanbul, Turkey, (e-mails: \{eylem.erdogan, nihat.kabaoglu\}@medeniyet.edu.tr).
}
\thanks{I. Altunbas is with the Department of Electronics and Communication Engineering, Istanbul Technical University, Istanbul, Turkey, (e-mail:
ibraltunbas@itu.edu.tr).}
\thanks{H. Yanikomeroglu is with the Department of Systems and
Computer Engineering, Carleton University, Ottawa, ON, Canada, (e-mail:
halim@sce.carleton.ca).} }

\IEEEtitleabstractindextext{\begin{abstract}Two emerging technologies, cognitive radio (CR) and free-space optical (FSO) communication, have created much interest both in academia and industry recently as they can fully utilize the spectrum while providing cost-efficient secure communication. In this article, motivated by the mounting interest in CR and FSO systems and by their ability to be rapidly deployed for civil and military applications, particularly in emergency situations, we propose a CR enabled radio frequency (RF)/FSO communication model for an aerial relay network. In the proposed model, CR enabled RF communication is employed for a ground-to-air channel to exploit the advantages of CR, including spectrum efficiency, multi-user connectivity, and spatial diversity. For an air-to-air channel, FSO communication is used, since the air-to-air path can provide perfect line-of-sight connectivity, which is vital for FSO systems. Finally, for an air-to-ground channel, a hybrid RF/FSO communication system is employed, where the RF communication functions as a backup for the FSO communication in the presence of adverse weather conditions. The proposed communication model is shown to be capable of fully utilizing the frequency spectrum, while effectively dealing with RF network problems of spectrum mobility and underutilization, especially for emergency conditions when multiple unmanned aerial vehicles (UAVs) are deployed.
\end{abstract}

}

\maketitle

\section{INTRODUCTION}

Over the past few years, there has been a mounting interest in the efficient use of the frequency spectrum as wireless data traffic has been increasing at a tremendous rate due to the proliferation of user equipment. The ever increasing demand for spectrum efficiency has trigged new spectrally efficient technologies including cognitive radio (CR) and free-space optical (FSO) communication. Among them, CR can utilize the spectrum access, and it offers notable advantages including energy efficiency, lower delay, and higher throughput. FSO communication can be used for high rate point-to-point communications over long distances. Compared to radio frequency (RF) communication, FSO systems operate over a much higher bandwidth (over $300$ GHz of unlicensed spectrum band); moreover, they have very narrow laser beams that can provide inherent security and robustness to interference.

CR technology can be realized by using three spectrum sharing paradigms: underlay, overlay and interweave. In the underlay approach, an unlicensed secondary user (SU) can use the same spectrum as a licensed primary user (PU) without creating any interference. Hence, energy efficient dynamic spectrum sharing can be enabled. In the overlay approach, the SU helps the PU by sharing its resources (e.g., time slot, power) to improve the communication capabilities of the PU. Finally, in the interweave approach, the SU uses the spectrum band of the PU when no active licensed users are in the system. In this technique, advanced spectrum sensing technologies aided with artificial intelligence are required to avoid unwanted interference. With the aid of these paradigms, CR can utilize the frequency spectrum efficiently \cite{Haykin}. 

FSO communication, on the other hand, can be described as the transmission of information by using light emitting diodes or laser beams with the aid of optical carriers over two stationary points operating in the near infrared (IR) band. In the current literature, FSO communication is usually enabled at the rooftops of the buildings to provide line-of-sight (LOS) connectivity \cite{Kedar}. Compared to its RF counterpart, FSO communication can boost the overall performance of wireless systems by enabling high data rates, lower delay, inherent security, and easy deployment. On account of this potential, FSO systems are expected to be used in many areas including campus connectivity, video surveillance, and monitoring. One of the major challenges in FSO systems is atmospheric turbulence induced fading, which is caused by adverse weather conditions. Specifically, fog, snow and rain can cause random fluctuations in the amplitude and phase of the optical transmission \cite{Uysal}. One of the solutions presented in the literature is to use FSO and RF communication together to exploit the advantages of optical communication, while mitigating the turbulence induced fading. In the so-called \quotes{mixed} RF/FSO communication, FSO systems can be employed with its RF counterpart in a dual-hop or multi-hop configuration with the aid of well-known relaying approaches. By contrast, in \quotes{hybrid} RF/FSO communication, RF transmission can be used as a backup for the FSO communication, and the system can switch between RF and FSO depending on the weather conditions \cite{Alzenad}.

Like CR and FSO systems, aerial relay (AR) aided communication is another promising technology for 5G+ networks. ARs can operate as flying base stations and provide perfect line-of-sight (LOS) connectivity to ground users and other ARs while being agile enough to maneuver and adjust their altitudes depending on weather conditions. ARs use amplify-and-forward (AF) and decode-and-forward relaying (DF) strategies to improve communication reliability and coverage \cite{Chen}. Owing to this potential, ARs have many fields of application, including traffic monitoring, search and rescue missions, remote sensing, and IoT scenarios where they can provide internet access to anywhere and anytime. Moreover, in future 5G+ networks, ARs are not only expected to be used for emergency situations, such as floods and earthquakes, but also for major political, sports, and music events. Indeed, \quotes{Project Loon} has already started to deploy ARs at altitudes of about $20$ km to provide greater connectivity to unserved and underserved communities \cite{Loon}.

 One of the the most important challenges in AR aided communication is the use of the frequency spectrum. The current spectrum is overcrowded, and the unlicensed frequency bands can create additional bandwidth and data rate problems for ARs. To solve these problems and to create a spectrally efficient AR network, this article proposes combining two emerging technologies, CR and FSO communication with an AR network, consisting of low-altitude unmanned aerial vehicle (UAV) relay nodes (URNs) and high-altitude platform station (HAPS) systems, as demonstrated in Fig. \ref{fig_1}. More precisely, we propose to use a CR enabled RF transmission for the ground-to-air (GtA) channel to create spectrally efficient multi-user connectivity, whereas FSO communication is used at the air-to-air (AtA) link, where a continuous LOS connection is available with the aid of acousto-optic and electro-optic FSO transceivers \cite{Fawaz}. Finally, for the air-to-ground (AtG) channel, hybrid RF/FSO communication is used, where the RF communication functions as a backup for the FSO communication, creating continuous connectivity without any performance loss.

\section{Background} 

\subsection{Types of ARs}

ARs can be classified into two groups on the basis of their altitudes: HAPS systems and URNs as they act as intermediate relay nodes for integrated access and backhaul by using AF and DF relaying strategies \cite{Irem}. URNs can be deployed at an altitude of up to $300$ m, whereas HAPS systems can be located in the stratosphere at $17$-$22$ km above the surface \cite{3GPP}. HAPS systems have wider coverage (about $200$-$300$ km in radius) and longer endurance (up to three months with the aid of solar energy panels). In HAPS communication, balloons, airships, airplanes and zeppelins may be used. Among them, balloons and zeppelins can hover at a fixed point that allows LOS conditions needed for the optical communication \cite{Fidler}. For example, in Project Loon, essential base station components were redesigned to be light enough to be carried by a balloon. Even though HAPS communications have attracted considerable interest in academia and industry, {HAPS systems can be vulnerable to the harsh weather conditions of stratosphere, where temperatures drop to $-90$$^{\circ}$C and wind speeds reach up to $100$ km per hour.} In addition, the use of the RF spectrum can create additional problems for HAPS nodes, especially in crowded metropolitan areas, and deployment time can be critical in the event of emergency situations.

URNs are flexible, highly maneuverable and can be deployed very quickly, unlike HAPS systems, making them suitable for emergency conditions, including earthquakes, floods, and search and rescue missions. URNs can be categorized as rotary-wing and fixed-wing URNs. Rotary-wing URNs have rotor blades and there are quadcopter (four rotors) and hexacopter (six rotors) designs; they can hover in a given area in the sky, which makes them preferable for LOS communication. Fixed-wing URNs are faster than rotary-wing URNs; however, they need a runway for takeoffs and landings. In addition, fixed-wing URNs can be exposed to significant Doppler spread as their relative velocities are very high. Another important challenge associated with URN communication is service time. URNs should be fully charged before being deployed, and they can only be used for a short time. 
\subsection{Channel Models}

\subsubsection{Ground-to-Air (GtA) Channel}

\paragraph{\textbf{Path Loss}}
{In RF communication, the path loss (PL) experienced by GtA or AtG paths can be expressed as $V(r) = \zeta   r^{\rho}$, where $\zeta$ is the PL at the reference point, $\rho$ is the PL exponent, $r = \sqrt{d^2 + h^2}$ is the distance between ground station and URNs, $d$ is the horizontal hop distance and $h$ shows the altitude of the URN \cite{Chen}.} In \cite{Chen}, three different types of deterministic PL models were proposed. In the Type A PL model, a quadcopter URN was used in the experiments and empirical results showed that the signal levels dropped quickly in the AtG or GtA channels compared to the AtA channel, which can be modeled with free-space PL. In the Type B PL model, URNs experienced shadowing and scattering effects due to man-made structures as they moved through urban areas. Finally, in the Type C PL model, the URN altitude was taken as $ 1\leq h \leq 120$ meters, and the PL for the AtG path was obtained by considering only the altitude of the URN.

\paragraph{\textbf{Fading Models}}

In AR aided RF communication, almost all existing fading models including Loo's model, Rayleigh, Rician, Nakagami-\emph{m} and Weibull have been used so far to model AtG or GtA paths \cite{Khuwaja}. More precisely, Loo's model is a composite channel model, in which the LOS component is modeled by a Log-normal distribution, whereas the multi-path components follow a Rician distribution. The Rayleigh channel, which is a special case of Rician and Nakagami-\emph{m}, is used for scattering environments. In the literature, Rayleigh distribution was adopted to the AR networks. Rician can be used to model LOS fluctuations of the channel, and is suitable for URNs and HAPS systems. In addition, Nakagami-\emph{m} and Weibull distributions can characterize HAPS communications \cite{Khuwaja}. 

It is noteworthy that, even though these classical fading models can be used for ARs including URNs and HAPS systems, they neglect the shadowing effect that can block the AtG or GtA communication due to huge mountains or manmade constructions as the control center or ground station may be located inside urban areas including metropolitan centers. For this reason, Shadowed-Rician and Shadowed-Rayleigh distributions can be more appropriate for modeling AtG and GtA communications. Specifically, Shadowed-Rayleigh can be used in highly shadowed places like shopping malls or skyscrapers where users do not have a LOS communication with HAPS systems or satellites. Likewise, in regions where LOS connectivity can be established, Shadowed-Rician model can be good fit for HAPS or satellite communications.

\begin{figure*}[t!]
\centering
\includegraphics[height=4.35in,width=7in]{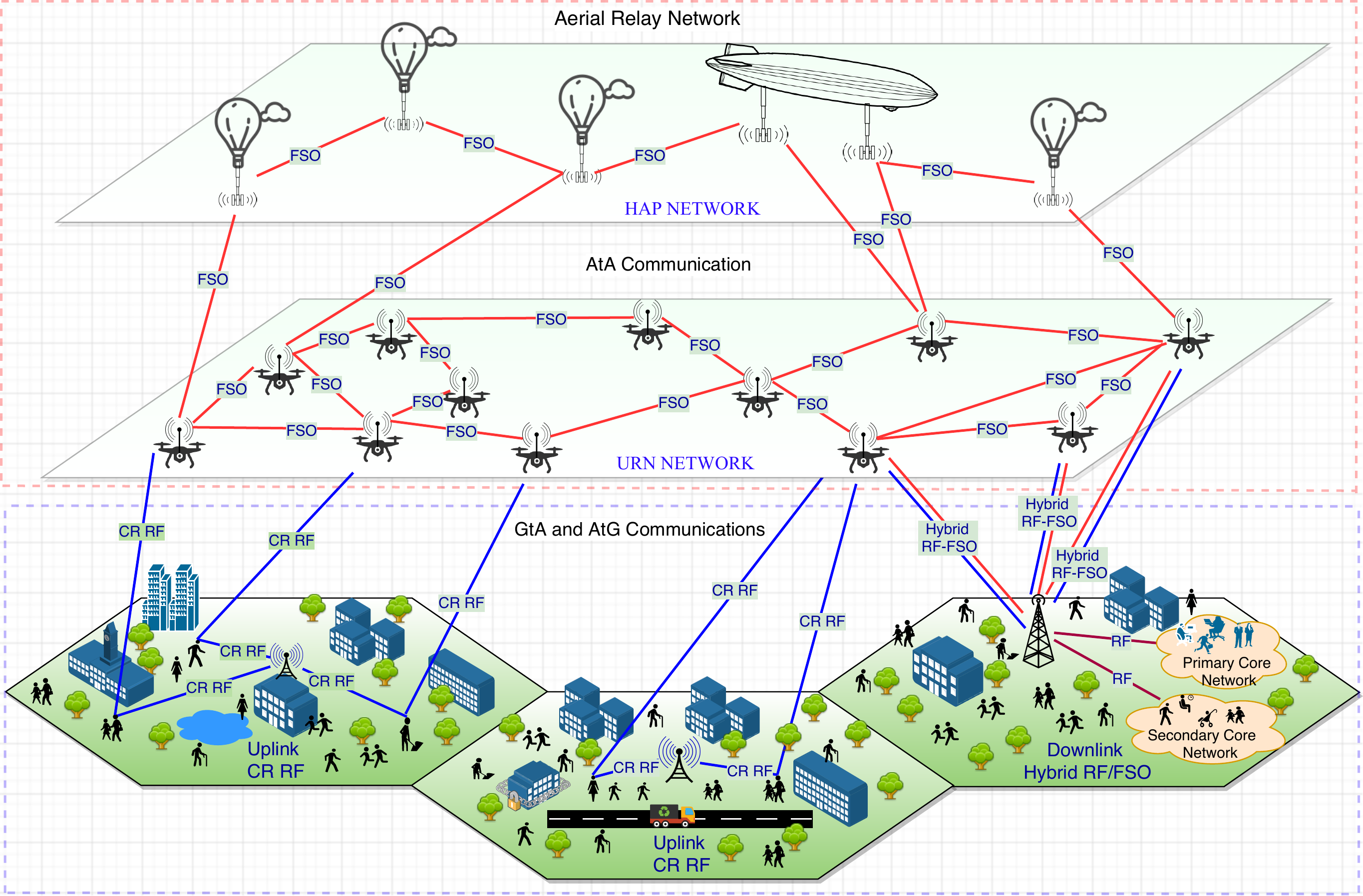}
\caption{Graphical Illustration of CR enabled RF/FSO communication model for Aerial Relay Network}
\label{fig_1}
\end{figure*}

\subsubsection{Air-to-Air (AtA) Channel}  

\paragraph{\textbf{Atmospheric Attenuation}}
Atmospheric attenuation, which can be caused by scattering and absorption during the propagation of the beam, is one of the major drawbacks in AtA FSO communication. {In general, atmospheric attenuation ($I_l$) is deterministic, and it can be modeled by using Beer-Lambert's law as $I_l = \exp(-\sigma L)$ \cite{Kaur}, where $L$ is the propagation distance between ARs and $\sigma$ is the attenuation coefficient which can be calculated with the aid of Kim's model \cite{Ghassemlooy}. Table \ref{Tab1} presents the attenuation coefficients and visibility parameters that can be used to determine the atmospheric attenuation for various fog conditions, when the wavelength is set to $\lambda = 1550$ nm.

\begin{table}[t]
	\caption{{Atmospheric attenuation and visibility parameters for different fog conditions at $\lambda = 1550$ nm}}
	\label{notations}
	\begin{center}
		\begin{tabular}{ l|l|l } 
			
			\hline
			Fog & Visibility (km)  & Attenuation coefficient (dB/km)   \\
			\hline \hline
			
			Dense &   $0.05$ & $339.62$  \\ 
			
			Thick &  $0.20$ & $84.90$  \\
			
			Moderate &  $0.50$ & $33.96$  \\
			
			Light &  $0.77$ & $16.67$   \\
			
			Thin  &  $1.90$ & $4.59$  \\

			\hline
		\end{tabular}
	\end{center}
	\label{Tab1}
\end{table}
\normalsize
}  

\paragraph{\textbf{Atmospheric Turbulence}}
In FSO communication, atmospheric turbulence ($I_a$) can be described as the random fluctuations of the received signal due to adverse weather conditions, including fog, snow, heavy rain, thunderstorms, etc. For AtA FSO communication, classical turbulence induced fading channels, including Log-normal and Gamma-Gamma have been introduced in the literature. Log-normal fading is appropriate for weak turbulence conditions, whereas Gamma-Gamma can be used in all turbulence regimes. Moreover, $\mathcal{K}$, Malaga  ($\mathcal{M}$) and Exponentiated Weibull (EW) fading can be used in AtA communication. Specifically, $\mathcal{K}$ distribution can precisely model strong turbulence induced fading, whereas $\mathcal{M}$ is a general distribution model that can cover classical turbulence channels, including Log-normal, Gamma-Gamma, and $\mathcal{K}$ \cite{Epple}, \cite{Dabiri1}. {Compared to its counterparts, EW can be adopted for the AtA communication as it can model weak-to-strong turbulence induced fading for various aperture sizes, including point-like receive apertures ($3$ mm), when aperture averaging takes place \cite{Barrios2}}. EW can be an important model for the AtA communication channel as URNs can have limited aperture sizes, and a point-like aperture can be employed in URNs \cite{Dabiri2}. 
 
\paragraph{\textbf{Geometric Loss}}
Geometric loss, which can be caused by the divergence of the beam between transmitter and receiver, is another important drawback in AtA communications. In geometric loss, the light spreads out as it moves from the transmitter to the photo detector, and the received power can be reduced . Another important cause of geometric loss is the angle of arrival fluctuations, which is caused by the position and orientation of ARs. {In the literature, geometric loss has been studied for UAV aided FSO front-haul channels, where deterministic and statistical geometric loss models were developed for fixed and variable URN positions \cite{Schober}. 

By considering atmospheric attenuation, atmospheric turbulence and geometrical loss ($I_g$), the aggregated channel model for FSO communication can be expressed as $I = I_aI_gI_l$ \cite{Wang2}.
}

\subsubsection{Air-to-Ground (AtG) Channel}  
In the AtG channel, a hybrid RF/FSO relaying procedure is employed. To this end, atmospheric attenuation and turbulence induced fading can be modeled similarly as given in AtA channel. In this model, geometric loss can be compensated as the destination ground station is in a fixed location. In the RF transmission, which is a backup for the FSO communication, GtA PL and fading models can be used. Specifically, Shadowed-Rician can be more appropriate in the RF communication as the ground station can be located in a highly shadowed area.

\section{Proposed System and Possible Configurations}

Consider a catastrophic emergency situation like a large earthquake where the entire communication system collapses. In such a scenario, a massive number of ARs, including URNs and HAPS systems, could be deployed in a short period of time to provide continuous communication connectivity and to help in the search rescue missions. Under such conditions, an important challenge, is the use of the frequency spectrum. So far, UAVs have been operating in the unlicensed band (IEEE L-Band, ISM band, etc.). However, these spectrum bands have become overcrowded as many other wireless systems, including WiMAX and Bluetooth, are using the same spectrum. To solve the spectrum scarcity problem, and to provide continuous connectivity, we propose combining CR enabled RF transmission and FSO communication for AR networks. The following subsections describe the proposed system and its advantages.

\subsection{System Model}


{{For the GtA channel, CR enabled RF communication is established to provide multi-user connectivity while enhancing the utilization of the spectrum}\footnote{{Please note that, it is almost impossible to use hybrid RF/FSO communication model in the GtA channel as user equipments are limited with an RF antenna and the user mobility can prohibit the LOS communication which is indispensable for the FSO communication.}}. In this model, the spectrum is utilized through spectrum sharing, where users/citizens behave like secondary nodes that would like to communicate with the other users in the system as depicted in Fig. \ref{fig_1}. In the underlay mode, SUs can communicate with the URNs by obeying the power constraints of the primary network, as long as the  SU related interference remains below the predetermined threshold.} In return, they may pay a roaming fee for being served by the primary network. In the overlay mode, multi-user connectivity can be established with the aid of SUs. More precisely, SUs can help the primary network to send a confidential message (governmental information) by creating a virtual multi-input single-output structure, and in return SUs can be served when the primary network is idle. The interweave mode, by contrast, is feasible when there are a limited number of PUs in the system. In CR enabled RF transmission, spectrum mobility and spectrum sensing are two important limiting factors as SUs need continuous and reliable connectivity with the AR networks. Spectrum sharing can be realized by using several different approaches, including energy detection, autocorrelation based detection, matched filter based detection, and Euclidean distance based detection. Among these techniques, energy detection can be preferable for the proposed setup as it depends only on the received signal energy. On the other hand, in spectrum mobility, the SUs should move from one licensed band to another when the aggregate interference exceeds the predefined threshold. The spectrum handoff decision is therefore very important in CR enabled RF transmission for AR networks to provide minimum delay \cite{Ivan}.

For the AtA channel, FSO communication can be employed between URNs and HAPS systems as they can provide perfect LOS connectivity. In this communication model, we assume that the ground signal is received by the hovering URNs, especially the ones that are closer to the ground terminal \cite{Dabiri3}. The URNs first filter out the DC components from the received signal, and then the electrical signal is converted to the optical signal by using subcarrier intensity modulation with proper biasing to ensure that the optical signal is positive. It is important to note that this conversion operation is simple and cheap as it only needs an optical transceiver, and almost all ARs can be equipped with this material. After the conversion is completed, URNs form a cooperation with other URNs or HAPS systems depending on the weather conditions or the other necessities. To do so, all URNs are equipped with an RF antenna and an optical transceiver, whereas HAPS systems can be employed with one or more optical transceivers as they have more available space. To exploit the benefits of the cooperation, different relaying protocols including serial, parallel, and bidirectional relaying can be used for the AtA channel \cite{Wang}. For example, serial relaying can be used between URNs when they are at similar heights, particularly in cases of search and rescue missions, when urgent action is needed. Parallel relaying can be used in the presence of harsh weather conditions when URNs can be aided by HAPS systems to increase the diversity and reliability of communication. On the other hand, as optical communication allows full duplex communication, bidirectional relaying can be used between different URNs and HAPS systems especially when urgent and important governmental messages are needed to be shared with survivors. In serial relaying, DF protocol can be preferable as AF relaying increases the total noise, whereas in parallel relaying, the AF mode can be beneficial for improving the overall performance of the system. In bidirectional relaying, on the other hand, AF and DF transmission can be used as both can reduce the total transmission time into two time slots. 


For the AtG channel, hybrid RF/FSO communication is employed where FSO is used as a primary transmission model as long as an LOS path is formed with the ground station. In the presence of adverse weather conditions or when the ground station is located in a shadowed urban area, RF communication can be used as a backup for the optical communication. {As mentioned above, all URNs are readily employed with RF antennas and FSO apertures to convert the electrical signal into optical in the GtA channel and the optical signal into electrical in the AtG channel}. Likewise, in the presence of heavy weather conditions, like dense fog, heavy snow or to improve reliability and diversity, URNs can transmit both electrical and optical information to the ground station. Hence, the overall spatial diversity can be doubled. In addition, more than one URN can be used for the AtG channel to further enhance the diversity gain, reliability, and the overall system performance. It is important to note that the ground station has the ability to convert the optical signal into an electrical signal by using a high speed photodiode that can deliver RF electrical signal output. Thereafter, it can combine all replicas of the RF received signals that are transmitted from different URNs by using appropriate combining methods, like maximum ratio combining, equal gain combining, or selection combining. Finally, the ground station delivers the primary and the secondary messages to the primary and secondary core networks.  

\subsection{Advantages}

\textbf{Battery Lifetime:}
As mentioned before, to convert the RF signal into an optical signal, all URNs should be outfitted with an RF antenna, a photodiode or an optical modulator. This procedure is energy efficient. In addition, FSO communication further improves battery life of ARs as laser beams (consisting of LEDs) spend less energy in the communication process. 

\textbf{Enhanced Performance:} 
The proposed system model can enhance overall performance of the AR networks as it exploits all the benefits of RF/FSO transmission. More precisely, a CR enabled GtA channel can provide diversity gain by allowing multi-user connectivity. The AtA channel can enable fast and ubiquitous communication by forming different cooperation strategies between URNs and HAPS systems, including serial, parallel, and bidirectional relaying. Finally, the AtG channel can provide diversity and enhanced performance as the information can be transmitted through RF and FSO antennas to the ground station. 

\textbf{Efficient Spectrum Usage:}
Both CR and FSO communication can utilize the frequency spectrum. More precisely, CR allows dynamic spectrum access in which both PUs and SUs can use the same spectrum, whereas FSO communication can remedy the problem of spectrum scarcity by allowing license free spectrum with Gigabit data rates. Therefore, FSO communication with its CR RF counterpart can be a spectrally efficient transmission model for AR networks. 

\textbf{Cost-Effective:} 
Dynamic spectrum access can bring a new perspective to the use of the wireless spectrum efficiently. This inevitably results in cheaper spectrum access. In addition, using simple FSO transceivers on URNs and HAPS systems, is inexpensive and can be easily implemented. 

\textbf{Inherent Security:}
AR networks can be more prone to security threats, especially since it is easy to tackle the information from a URN that is hovering at a specific altitude. However, due to the specific nature of FSO systems, the eavesdropper can not tackle the information from an optical beam. Hence, a secure FSO communication can be established between AR nodes.

\begin{figure}[t!]
\centering
\includegraphics[height=5in,width=3.5in]{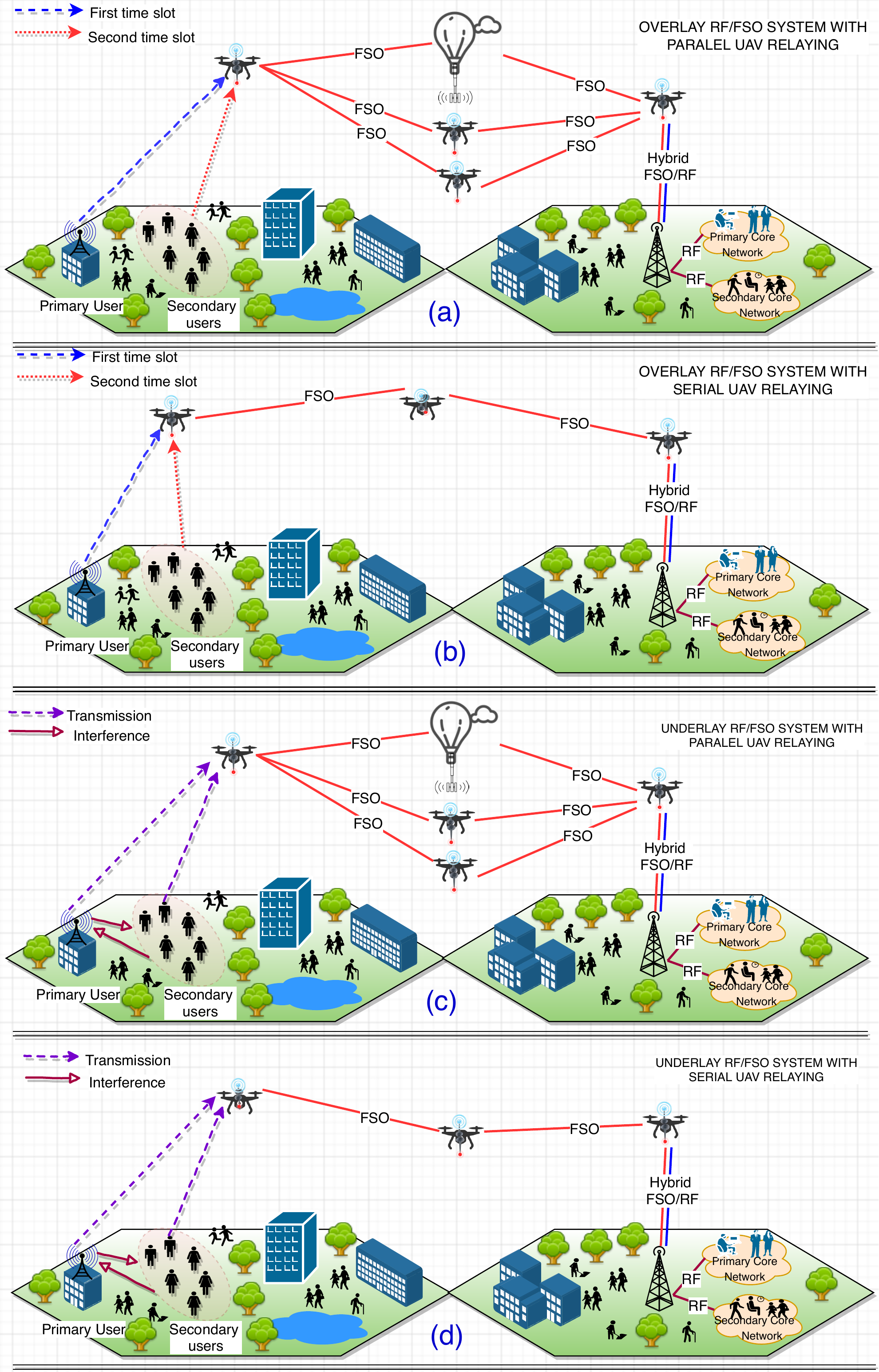}
\caption{a) Overlay RF/FSO transmission with parallel relaying b) Overlay RF/FSO transmission with serial relaying c) Underlay RF/FSO transmission with parallel relaying d) Underlay RF/FSO transmission with serial relaying.}
\label{fig_2}
\end{figure}


\section{Performance Evaluation}
{
In this section, we consider four practical CR RF/FSO relaying modes including underlay/overlay serial and parallel DF relaying\footnote{{In this section DF mode is used to make a fair comparison of serial and parallel relaying schemes.}} scenarios as depicted in Fig. 2. To quantify the performance of the proposed schemes, we present the statistical properties of the signal-to-noise ratio (SNR) for GtA, AtA and AtG channels, and provide outage probability analysis. Thereafter, simulation parameters are introduced, numerical examples are illustrated, and important design guidelines are outlined.

\subsection{Statistical Properties of the SNR}
\subsubsection{GtA Communication} 
As depicted in Fig. 2, CR enabled RF transmission is adopted in the GtA channel, where both overlay and underlay schemes are used depending on the availability of the primary network. In the overlay mode{\footnote{{In the overlay mode, we assume that the primary network is idle, so that the secondary users can communicate with the AR network without any power limitations.}}}, the signal-to-noise ratio (SNR) of the GtA channel can be expressed as 
\begin{align}
	\gamma_\text{GtA}^{\text{o}} = \frac{P_s|g|^2}{P_nV(r)},
	\label{EQN:1}
\end{align}
where $P_s$ and $P_n$ are the transmit and noise powers,  $V(r)$ is the path-loss variable defined in Section II, and  $|g|$ is the channel magnitude, which is distributed with Nakagami-\emph{m} fading. In the underlay mode, the SNR of the GtA channel can be written as \cite{Eylem}
\begin{align}
	\gamma_\text{GtA}^{\text{u}} = \frac{P_s}{P_nV(r)}\times\frac{|g|^2}{E[|h_I|^2]},
	\label{EQN:2}
\end{align}
where $E[\cdot]$ shows the expectation operator and $h_I$ is the secondary-primary interference term which can be obtained through mean-value power allocation model presented in \cite[Sect. 4.4]{Eylem}. In the presence of primary-secondary interference, the effective SNR of the GtA channel can be expressed as 
\begin{align}
	\gamma_\text{GtA}^{\text{u,ef}} = \frac{\gamma_\text{GtA}^{\text{u}}}{1+\mathcal{I}_P},
	\label{EQN:3}
\end{align}
where $\mathcal{I}_P$ shows the fixed primary-secondary interference SNR. For underlay and overlay modes, the cumulative distribution function (CDF) of the SNR can be expressed as 
\begin{align}
F_{\gamma_\text{GtA}}(\gamma) =  1-e^{-\frac{m\gamma}{\bar{\gamma}_\text{GtA}}} \sum_{v=0}^{m-1}\Big(\frac{m\gamma}{\bar{\gamma}_\text{GtA}}\Big)^v\frac{1}{v!},
\label{EQN:4}
\end{align}
where $\bar{\gamma}_\text{GtA} = \bar{\gamma}_\text{GtA}^{\text{o}} = \frac{P_s}{P_nV(r)}$ for the overlay mode and $\bar{\gamma}_\text{GtA} = \bar{\gamma}_\text{GtA}^{\text{u},\text{ef}} = \frac{\frac{P_s}{P_nV(r)E[|h_I|^2]}}{1+\mathcal{I}_P}$ for the underlay scheme. 

\vspace{0.1cm}
\subsubsection{AtA Communication}
In AtA communication, FSO transmission is adopted between ARs where communication is affected both from atmospheric attenuation and turbulence induced fading which is distributed with EW fading. In this case, the aggregated channel model can be written as{\footnote{{In the AR aided communication, geometric losses can be compensated by using tracking enabled electro-optic and acousto-optic FSO transceivers \cite{Fawaz}. Thereby, we neglect the geometric loss in this work}.}} $I=I_aI_l$, and the SNR of AtA channel can be expressed as 
\begin{align}
\gamma_{\text{AtA}} =  \frac{P_s}{P_n}I_a^2|I_l|^2,
	\label{EQN:5}
\end{align}
and the CDF of SNR for AtA channel can be obtained as \cite{Barrios2}
\begin{align}
F_{\gamma_{\text{AtA}}}(\gamma) = \Bigg(1-\exp\Bigg[-\Bigg(\frac{\gamma}{\eta^2\bar{\gamma}_{\text{AtA}}}\Bigg)^{\beta/2}\Bigg]\Bigg)^{\alpha}, 
	\label{EQN:6}
\end{align}
where $\bar{\gamma}_{\text{AtA}} = \frac{P_s}{P_n}I_a^2E[|I_l|^2]$ is the average SNR of the AtA channel, $\alpha$ and $\beta$ are the shape parameters and $\eta$ is the scale parameter of the EW fading channel. 

\subsubsection{AtG Communication}
In AtG communication, hybrid RF/FSO communication model is used where FSO is adopted as the primary transmission model as long as an LOS path is formed with the ground station. Therefore, the CDF of AtG channel can be obtained very similar to (\ref{EQN:6}) 

\subsection{Outage Probability Analysis}
Outage probability ($P_{\text{out}}$) is an important performance metric that can assess the overall performance of a wireless system. It can be defined as the probability of SNR that will fall below a predetermined threshold ($\gamma_{\text{th}}$) for acceptable communication quality. Mathematically, it can be formulated as 
\begin{align}
	P_{\text{out}} = \Pr\Big(\gamma_{e2e}<\gamma_{\text{th}}\Big),
	\label{EQN:7}
\end{align} 
where $\gamma_{e2e}$ is the end-to-end (e2e) SNR. In this section, the outage probability (OP) analysis is pursued for DF serial and parallel relaying networks.

For a DF serial relaying system, the e2e SNR can be expressed as 
\begin{align}
\gamma_{e2e}^{\text{S}} = \min\Big(\gamma_1,\gamma_2,\ldots,\gamma_N\Big), 	
	\label{EQN:8}
\end{align}
where $N$ denotes the number of hops and $\gamma_j$, $j\in\{1,N\}$ shows the SNR of the $j$-th hop. Similarly, for a parallel dual-hop DF relaying, the e2e SNR can be expressed as 
\begin{align}
\gamma_{e2e}^{\text{P}} = \max_{1\leq k\leq K}\Big(\min\Big(\gamma_{1,k},\gamma_{2,k}\Big)\Big),
	\label{EQN:9}
\end{align}
where $\gamma_{1,k}$ and $\gamma_{2,k}$ shows the SNRs of the first and second hops, and $K$ denotes the number of relays that are available for communication. By substituting \eqref{EQN:8} and \eqref{EQN:9} into \eqref{EQN:7}, the outage probabilities for serial and parallel DF relaying schemes can be obtained as \cite{Wang}
\small
\begin{align}
P_{\text{out}}\hspace{-0.05cm} = \hspace{-0.1cm}
\begin{cases}
1 - \prod_{j=1}^{N}(1-F_{\gamma_j}(\gamma_{\text{th}})), & \hspace{-0.2cm} \text{Serial R.,}\\
\prod_{k=1}^{K}\Big(1-(1-F_{\gamma_{1,k}}(\gamma_{\text{th}}))(1-F_{\gamma_{2,k}}(\gamma_{\text{th}}))\Big), & \hspace{-0.2cm}\text{Parallel R.,}     
\end{cases}
\label{EQN:10}
\end{align}
\normalsize 
where $F_{\gamma_{z,k}}(\gamma_{\text{th}})$, $z\in\{1,2\}$ and $F_{\gamma_j}(\gamma_{\text{th}})$ denotes CDF of the SNR per hop. By substituting (\ref{EQN:4}) and (\ref{EQN:6}) into  (\ref{EQN:10}), and after several manipulations, the OPs for underlay/overlay serial and parallel relaying can be obtained.

\subsection{Simulation Setup}
In this section, we provide the simulation parameters that are used to illustrate the practical CR RF/FSO relaying scenarios, which are depicted in Fig. 2. In the simulations, we assume that the FSO communication is affected both from turbulence induced fading and atmospheric attenuation, whereas the RF communication is impaired from path-loss and multi-path fading. Furthermore, in parallel relaying, the HAPS aided communication link can be affected from the atmospheric attenuation more than URNs, by contrast, URN to URN communication can experience strong turbulence effects as they are deployed at a lower altitude. Based on this assumption, the average SNRs are assumed equal. In the calculation, transmit power $P_t$ and noise power $P_n$ are set to $P_t = 32$ dBm, and $P_n = -100$ dBm, respectively. Moreover, the altitude of the HAPS and URNs are set to $h_s = 19000$ m, and $h_s = 200$ m, respectively, and the SNR threshold is given as $\gamma_{\text{th}} = 3$ dB for acceptable communication quality. In the overlay mode, primary-secondary interference can be neglected as it uses precise spectrum sensing techniques, whereas in the underlay mode, fixed $\mathcal{I}_P$ values varying from $0$ dB to $5$ dB, are used. Moreover, in FSO communication, Hufnagel-Valley day model is used to obtain the refractive-index structure parameter \cite{Barrios2}, and thin fog weather condition ($\sigma = 4.59$) is used to obtain atmospheric attenuation variable $I_l$. On the contrary, in RF communication, Type A PL model \cite{Chen} is used and the PL variable is obtained by setting $\alpha = 2.32$, $\beta = \Big(\frac{2\pi f_c}{c}\Big)^2$, $f_c = 2$ GHz carrier frequency, and Nakagami-\emph{m} fading parameter is set to $m=4$. Unless otherwise stated, Table II presents all simulation parameters.

\begin{table}[t]
\caption{LIST OF PARAMETERS AND VALUES}
\label{notations}
\begin{center}
\begin{tabular}{    l| l } 

Parameter & Value \\
\hline \hline

Optical wavelength ($\lambda$)  & $1550$ nm   \\ 

SNR threshold ($\gamma_{\text{th}}$) & $3$ dB \\

Nakagami-\emph{m} severity parameter ($m$)  & 4 \\

Optical to electrical conversion ratio   & $1$  \\

Primary-secondary interference SNR ($\mathcal{I}_P$) & $0$-$5$ dB \\

Horizontal hop distance ($d$)  & $2500$ m \\ 

URN altitude ($h_u$) & $200$ m \\

HAPS altitude ($h_s$) & $19$ km \\

Carrier frequency ($f_c$) & $2$ GHz \\

Attenuation coefficient ($\sigma$) & $4.5859$ \\

Wind speed ($v_g$) & $21$ m/s \\

\hline
\end{tabular}
\end{center}
\end{table}


\subsection{Numerical Illustrations}

Fig. \ref{fig_3} shows the OP as a function of horizontal hop distance ($d$) for the proposed overlay/underlay serial and parallel relaying systems, when the primary-secondary interference is set to $\mathcal{I}_P = 5$ dB. As we can see, parallel relaying outperforms serial relaying both in underlay and overlay mode, as it forms a cooperation to enhance the system performance. We also note that, underlay mode performs worse than its overlay counterpart due to high $\mathcal{I}_P$ levels. Furthermore, OP monotonically increases when $d$ goes from $1500$ m to $2600$ m. Then, due to huge atmospheric attenuation and PL effects, the OP performance worsens and when $d$ reaches to $3300$ m, the OP saturates and reaches to a outage ceiling where $P_{\text{out}} \approx 1$. Therefore, as can be observed from the figure, the optimum OP performance can be obtained when $d < 2000$ m, where $P_{\text{out}} < 10^{-4}$ for the overlay mode. 

\begin{figure}[t!]
\centering
\includegraphics[width=3.5in]{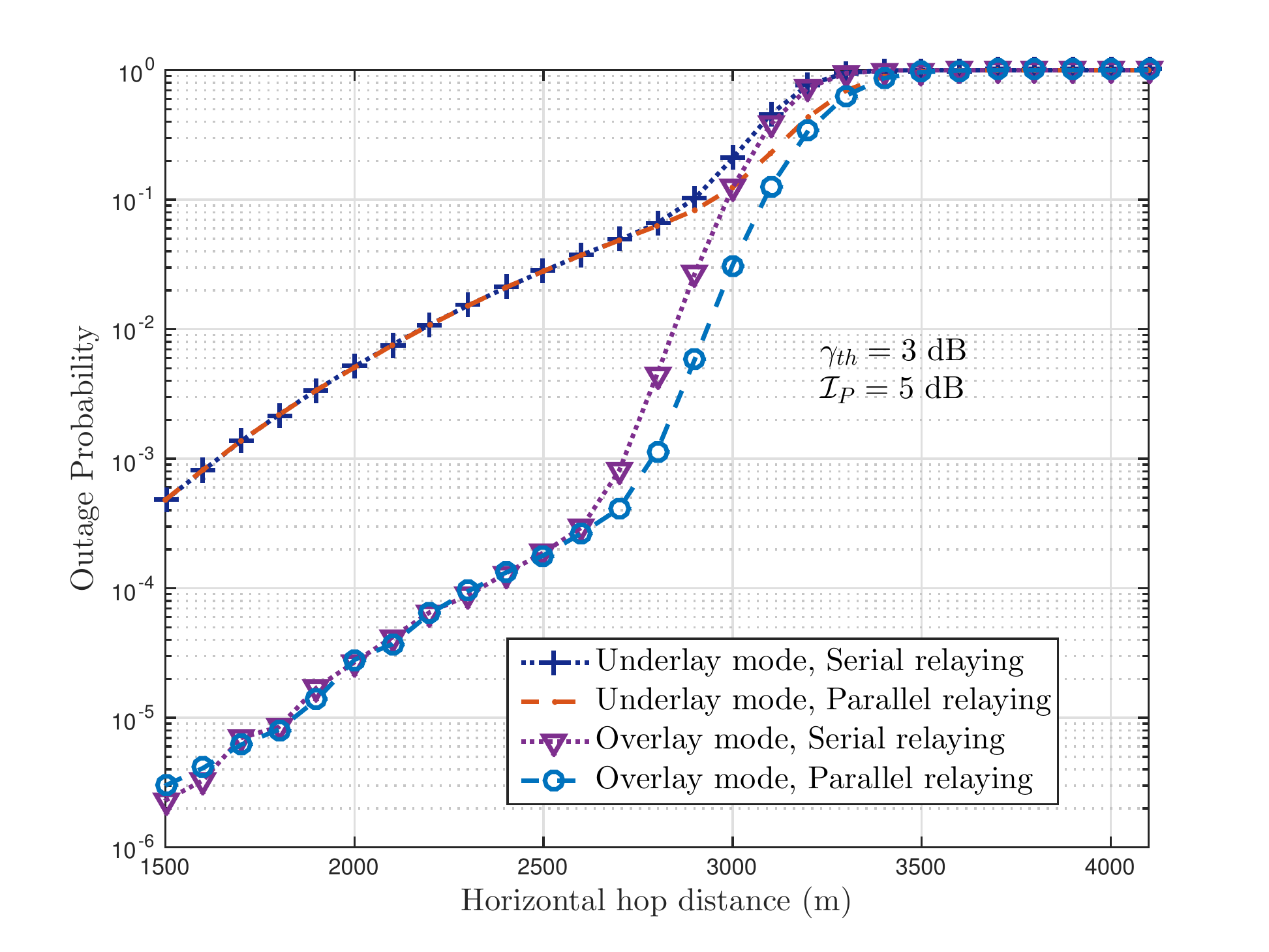}
\vspace{-0.15cm}
\caption{{Outage probability versus horizontal hop distance of overlay/underlay serial and parallel relay networks.}}
\label{fig_3}
\end{figure}
\begin{figure}[t!]
\centering
\includegraphics[width=3.5in]{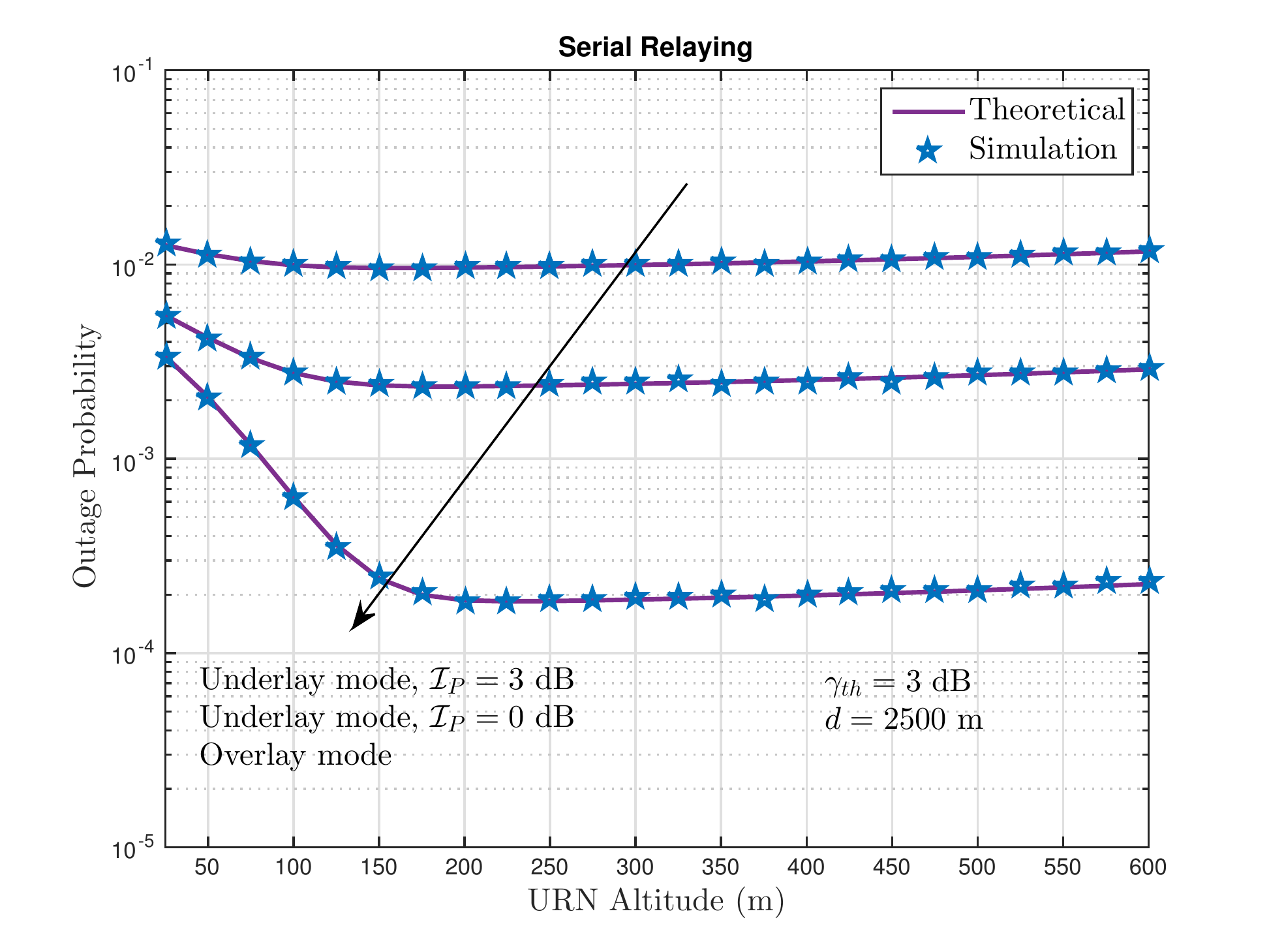}
\caption{{ Outage probability versus URN altitude of overlay and underlay serial relaying.}}
\label{fig_4}
\end{figure}

Fig. \ref{fig_4} illustrates the OP performance of underlay and overlay modes for serial relaying as a function of the URN altitude ($h_u$) for different $\mathcal{I}_P$ levels, when $d = 2500$ m. As can be observed, the theoretical outage probability results, which are shown with solid lines, are in good agreement with the marker symbols, which are generated by the simulations. Interestingly, the outage performance of the proposed scheme enhances with respect to the $h_u$ as the refractive-index structure parameter is directly related with URN altitude. More precisely, URN to URN communication can be less affected from the atmospheric turbulence induced fading when they are deployed at higher altitudes. On the contrary, when $h_u$ is higher than $200$ m, the OP performance of the proposed scheme worsens slightly due to huge PL and atmospheric attenuation effects. Thereby, placing the URNs at about $200$ m can provide optimum OP performance for the proposed scheme. Finally, it can be observed from the figure that, as $\mathcal{I}_P$ increases, the OP performance worsens in the underlay scheme. Therefore, the overlay scheme can be preferable for the GtA channel especially when the primary network is idle.

}

\subsection{Design Guidelines}

In this subsection, we shed light on the design of practical CR enabled RF/FSO AR networks by providing some important design guidelines. 

\begin{itemize}
	\item As can be seen in Fig. 3 and 4, the overlay method can be preferable in GtA communication as it allows multi-user access to the URNs through the primary spectrum without causing any interference. 
	\item In the AtA channel, DF relaying can be preferable when serial relaying is used as it does not enhance the noise level like AF relaying. On the other hand, when parallel or bidirectional relaying is used, the AF protocol can be 	preferable for enhancing the system performance.
	\item To provide a perfect LOS connection, URNs can be positioned close enough to each other in the presence of adverse weather conditions. In addition, HAPS systems can aid URNs in parallel relaying mode. 
	\item In the AtG channel, multiple URNs can be used to serve as a backhaul. Moreover, to provide reliable communication, both RF and FSO communications can be used simultaneously, depending on the urgency of the status or weather conditions. 
	\item Using FSO communication for the AtA channel can further improve the battery life of the URNs and HAPS systems. Moreover, FSO transceivers are cheap, easy to deploy, and consume less battery power as they use simple laser beams. 
	\item As all URNs are fitted with both RF and FSO antennas, hybrid RF/FSO communication model can be used for all URNs as needed. It is worth noting that hybrid RF/FSO communication uses more battery power and thus decreases the battery life of ARs quicker.
\end{itemize}

\section{Conclusion}

This article has proposed a new transmission model for aerial relay networks by combining spectrally efficient cognitive radio with FSO communication. More precisely, we showed how employing cognitive radio enabled RF communication for the ground-to-air channel, FSO communication for the air-to-air path, and a hybrid RF/FSO communication system for the air-to-ground channel can leverage the advantages of RF/FSO transmission, while effectively addressing the problem of spectrum scarcity. As we observed, the proposed setup utilizes the frequency spectrum, enhances data rates, and provides inherent security.

\end{document}